\documentclass[a4paper,11pt]{article}
\usepackage{url}
\usepackage{pos}
\usepackage{caption}
\usepackage{subcaption}

\title{Positioning system for Baikal-GVD}
\author[a]{V.A.~Allakhverdyan}
\author*[b]{A.D.~Avrorin}
\author[b]{A.V.~Avrorin}
\author[b]{V.M.~Aynutdinov}
\author[c]{R.~Bannasch}
\author[d]{Z.~Barda\v{c}ov\'{a}}
\author[a]{I.A.~Belolaptikov}
\author[a]{I.V.~Borina}
\author[a,1]{V.B.~Brudanin}
\author[e]{N.M.~Budnev}
\author[a]{V.Y.~Dik}
\author[b]{G.V.~Domogatsky}
\author[b]{A.A.~Doroshenko}
\author[a,d]{R.~Dvornick\'{y}}
\author[e]{A.N.~Dyachok}
\author[b]{Zh.-A.M.~Dzhilkibaev}
\author[d]{E.~Eckerov\'{a}}
\author[a]{T.V.~Elzhov}
\author[f]{L.~Fajt}
\author[g,1]{S.V.~Fialkovski}
\author[e]{A.R.~Gafarov}
\author[b]{K.V.~Golubkov}
\author[a]{N.S.~Gorshkov}
\author[e]{T.I.~Gress}
\author[a]{M.S.~Katulin}
\author[c]{K.G.~Kebkal}
\author[c]{O.G.~Kebkal}
\author[a]{E.V.~Khramov}
\author[a]{M.M.~Kolbin}
\author[a]{K.V.~Konischev}
\author[h]{K.A.~Kopa\'{n}ski}
\author[a]{A.V.~Korobchenko}
\author[b]{A.P.~Koshechkin}
\author[i]{V.A.~Kozhin}
\author[a]{M.V.~Kruglov}
\author[b]{M.K.~Kryukov}
\author[g]{V.F.~Kulepov}
\author[h]{Pa.~Malecki}
\author[a]{Y.M.~Malyshkin}
\author[b]{M.B.~Milenin}
\author[e]{R.R.~Mirgazov}
\author[a]{D.V.~Naumov}
\author[a]{V.~Nazari}
\author[h]{W.~Noga}
\author[b]{D.P.~Petukhov}
\author[a]{E.N.~Pliskovsky}
\author[j]{M.I.~Rozanov}
\author[a]{V.D.~Rushay}
\author[e]{E.V.~Ryabov}
\author[b]{G.B.~Safronov}
\author[a]{B.A.~Shaybonov}
\author[b]{M.D.~Shelepov}
\author[a,d,f]{F.~\v{S}imkovic}
\author[a]{A.E. Sirenko}
\author[i]{A.V.~Skurikhin}
\author[a]{A.G.~Solovjev}
\author[a]{M.N.~Sorokovikov}
\author[f]{I.~\v{S}tekl}
\author[b]{A.P.~Stromakov}
\author[a]{E.O.~Sushenok}
\author[b]{O.V.~Suvorova}
\author[e]{V.A.~Tabolenko}
\author[e]{B.A.~Tarashansky}
\author[a]{Y.V.~Yablokova}
\author[c]{S.A.~Yakovlev}
\author[b]{D.N.~Zaborov}

\affiliation[a]{Joint Institute for Nuclear Research, Dubna, Russia}
\affiliation[b]{Institute for Nuclear Research, Russian Academy of Sciences, Moscow, Russia}
\affiliation[c]{EvoLogics GmbH, Berlin, Germany}
\affiliation[d]{Comenius University, Bratislava, Slovakia}
\affiliation[e]{Irkutsk State University, Irkutsk, Russia}
\affiliation[f]{Czech Technical University in Prague, Prague, Czech Republic}
\affiliation[g]{Nizhny Novgorod State Technical University, Nizhny Novgorod, Russia}
\affiliation[h]{Institute of Nuclear Physics of Polish Academy of Sciences (IFJ~PAN), Krak\'{o}w, Poland}
\affiliation[i]{Skobeltsyn Institute of Nuclear Physics, Moscow State University, Moscow, Russia}
\affiliation[j]{St.~Petersburg State Marine Technical University, St.Petersburg, Russia}

\note{Deceased.}

\emailAdd{avrorin@inr.ru}

\abstract{
  Baikal-GVD is a kilometer scale neutrino telescope currently under construction in Lake Baikal.
  Due to water currents in Lake Baikal, individual photomultiplier housings are mobile and can drift away from their initial position.
  In order to accurately determine the coordinates of the photomultipliers, the telescope is equipped with an acoustic positioning system.
  The system consists of a network of acoustic modems, installed along the telescope strings and uses acoustic trilateration to determine the coordinates of individual modems.
  This contribution discusses the current state of the positioning in Baikal-GVD, including the recent upgrade to the acoustic modem polling algorithm.
}

\FullConference{%
  *** 37th International Cosmic Ray Conference (ICRC2021), ***\\
  *** 12-23 July 2021 ***\\
  *** Berlin, Germany - Online ***
  }

\begin{document}

  \maketitle

  \section{Introduction}

\begin{figure}
    \centering
    \begin{minipage}{0.45\textwidth}
      \centering
      \includegraphics[height=5.5cm]{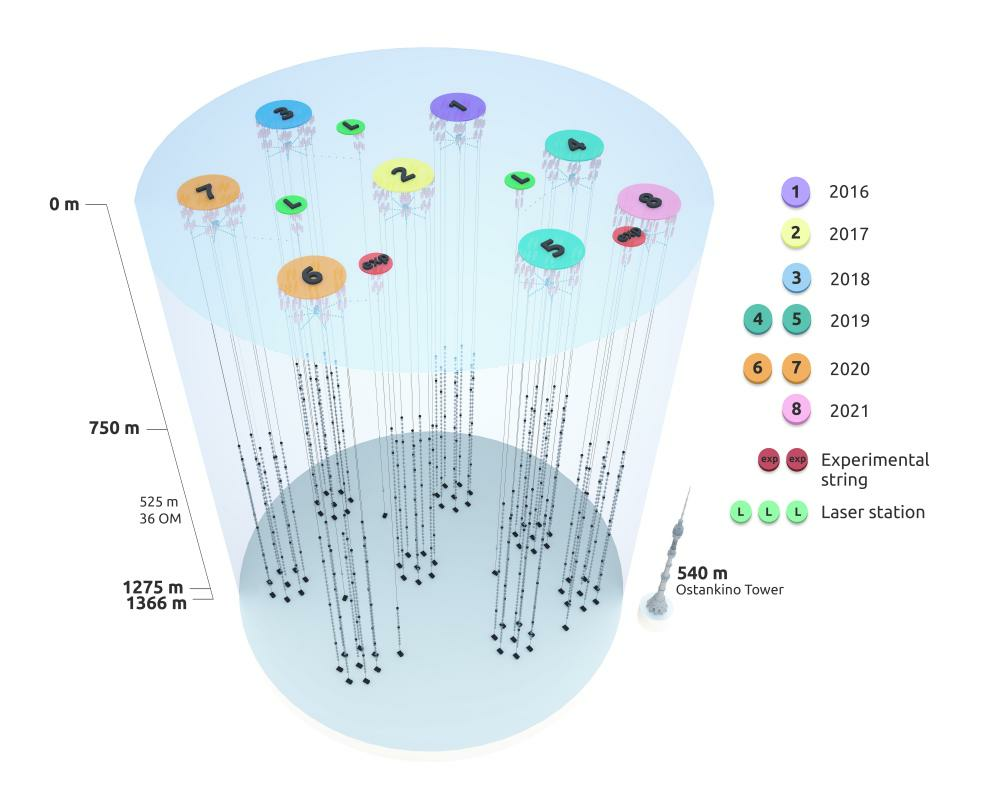}
      \caption{Baikal-GVD configuration in 2021}
      \label{fig:gvd2021}
    \end{minipage}
    \begin{minipage}{0.45\textwidth}
      \centering
      \includegraphics[height=5.92cm]{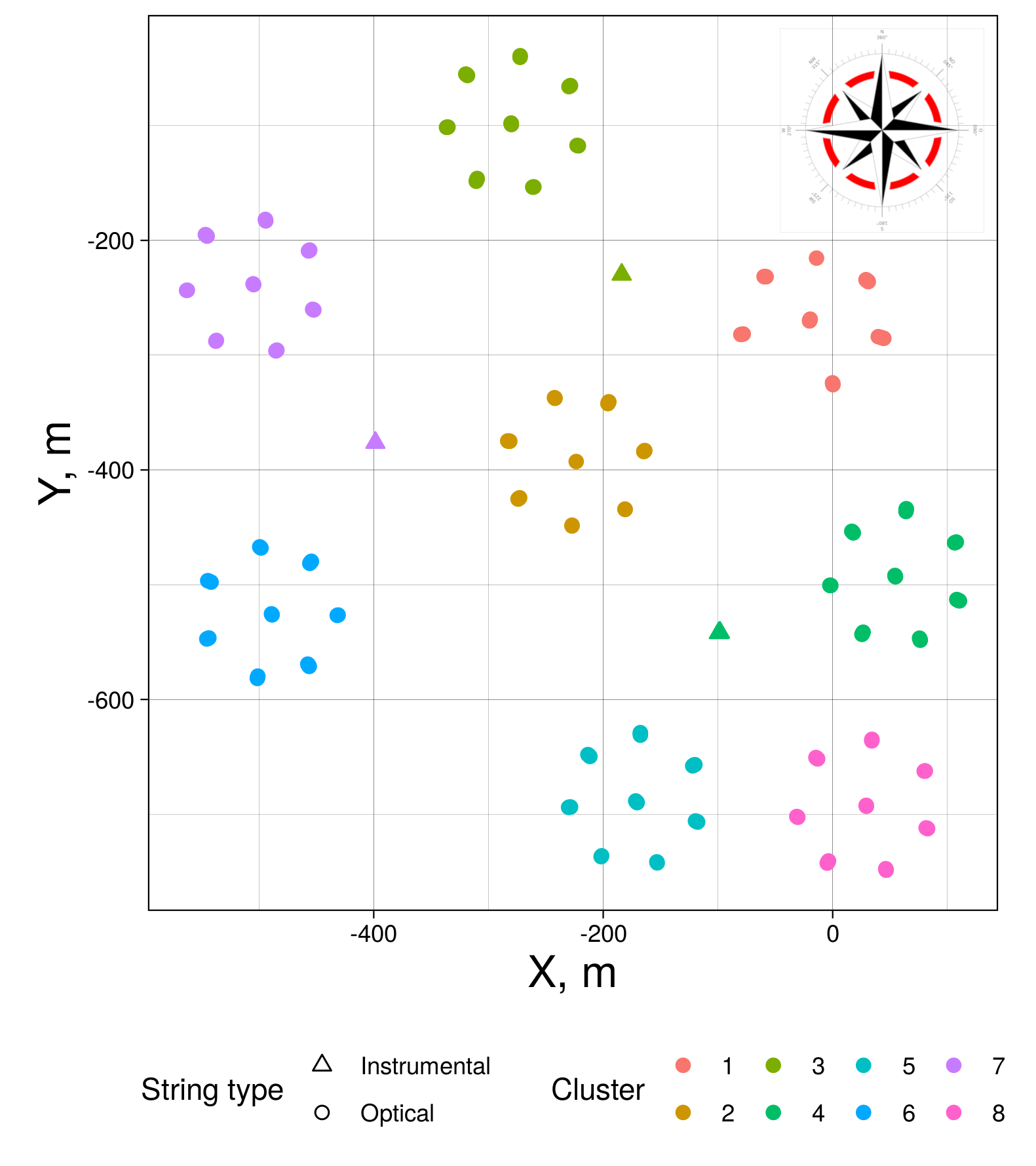}
      \caption{Reconstructed planar AM coordinates in GVD 2021. X is eastbound, Y is northbound.}
      \label{fig:xy_2021}
    \end{minipage}
  \end{figure}

  Baikal-GVD \cite{gvd, gvd2021} is a large scale underwater neutrino telescope currently under construction in Lake Baikal.
  The telescope operates by registering Cherenkov radiation from charged particles produced in neutrino interactions with Baikal water with a set of submerged photomultipliers and reconstructing neutrino direction and energy from the timing of PMT hits and their charge deposition.
  Individual photomultipliers are mounted in transparent spherical housings (called optical modules, or OMs), which are installed on flexible cables at the depths between 750 and 1275 meters.
  The bottom end of such cable is attached to an anchor installed on a lakebed, while the top end is suspended by subsurface buoys, 30 meters below the lake surface.
  There are 36 OMs installed on each cable with 15 meter intervals, comprising a string.
  The strings are organized in clusters, with 8 strings per cluster, totalling 288 OMs per cluster.
  7 strings in cluster are installed in the vertices of a regular heptagon with the radius of 60 m, the 8th string is installed in its center.
  Currently, 8 clusters are installed as demonstrated on Figure \ref{fig:gvd2021}.

  Due to the currents in Lake Baikal, OMs can drift away from their initial positions, introducing uncertainty to PMT hit timing - one meter of PMT drift introduces an about 4 ns error in hit timing uncertainty.
  The operation of the Baikal-GVD since 2016 has shown that some PMTs may drift beyond 50 meters from their initial positions. 
  To determine OM coordinates at the moment of the event detection, GVD is equipped with an acosutic positioning system (APS).

  \section{The Baikal-GVD positioning system}
  We use a network of acoustic modems (AMs) installed along the strings to determine geometry of the detector.
  Acoustic communication between AMs is described in \cite{Kebkal}.
  Acoustic modems installed near string anchors (nodes) are used to trilaterate coordinates of modems installed along the strings (beacons).
  Beacon coordinates are reconstructed online and the acoustic distances are measured every 100-200 seconds.
  %Acoustic distance measurement precision as estimated during winter expedition 2020: several centimeters at the distance of approximately 1500 meters.
  Coordinates of the mounted components, including OMs, are linearly interpolated from beacon coordinates.
  %OM positioning precision depends on season and elevation, but has been estimated to be $\sim 15$ cm.
  OM rotation around the string is not currently taken into account.

  Planar beacon coordinates for GVD 2021 configuration are presented on \ref{fig:xy_2021}.

  \begin{figure}
    \centering
    \begin{minipage}{0.45\textwidth}
      \centering
      \includegraphics[height=5.5cm]{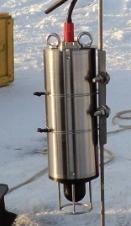}
      \caption{EvoLogics S2CR 42/65 acoustic modem}
      \label{fig:am_view}
    \end{minipage}
    \begin{minipage}{0.45\textwidth}
      \centering
      \includegraphics[height=5.5cm]{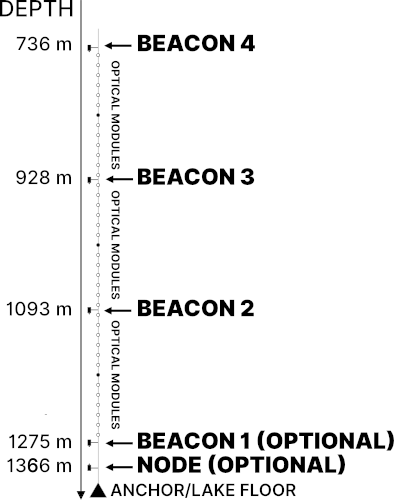}
      \caption{Acoustic modem layout on a GVD string}
      \label{fig:am_layout}
    \end{minipage}
  \end{figure}

  AM layout on a GVD string is presented on Figure \ref{fig:am_layout}.
  There are currently 280 AMs installed on GVD strings, including 35 nodes.
  An acoustic modem mounted on a string shortly before being submerged is presented on Figure \ref{fig:am_view}.
  The APS is operated independently from the GVD trigger system and data acquisition.
  AMs are polled by the shore software via RS-485 lines for inter-AM acoustic distances, which are then used to reconstruct beacon coordinates. 
  Reconstructed data is stored at the shore and is synchronized with the central APS database in JINR every 2 minutes.
 
  \section{Performance}

  \subsection{APS polling}
\begin{figure}[h]
    \centering
    \begin{subfigure}[t]{0.45\textwidth}
      \centering
      \includegraphics[width=\textwidth]{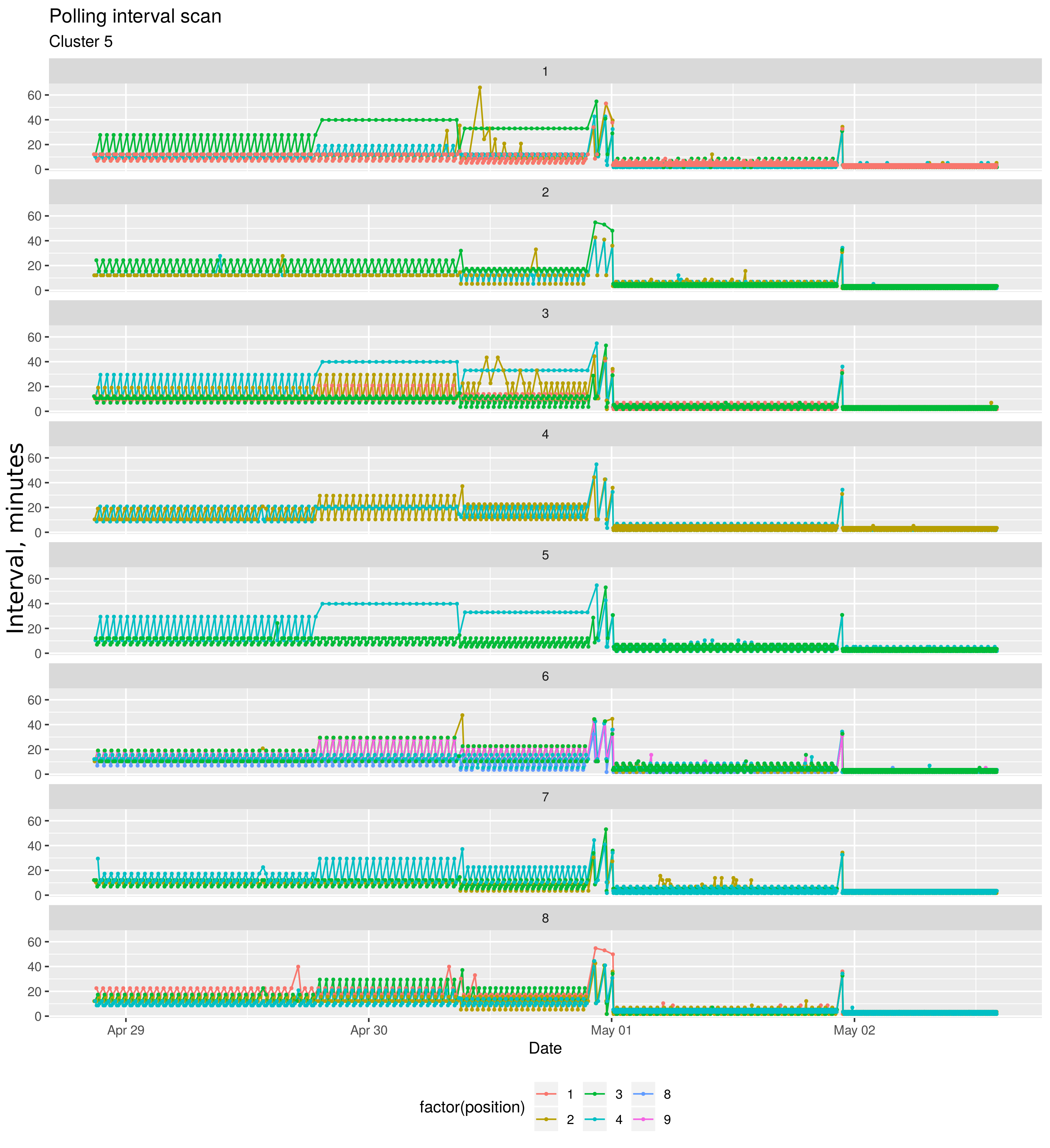}
      \caption{Polling intervals for beacons on cluster 5}
      \label{fig:ierr_1}
    \end{subfigure}
    \hfill
    \begin{subfigure}[t]{0.45\textwidth}
      \centering
      \includegraphics[width=\textwidth]{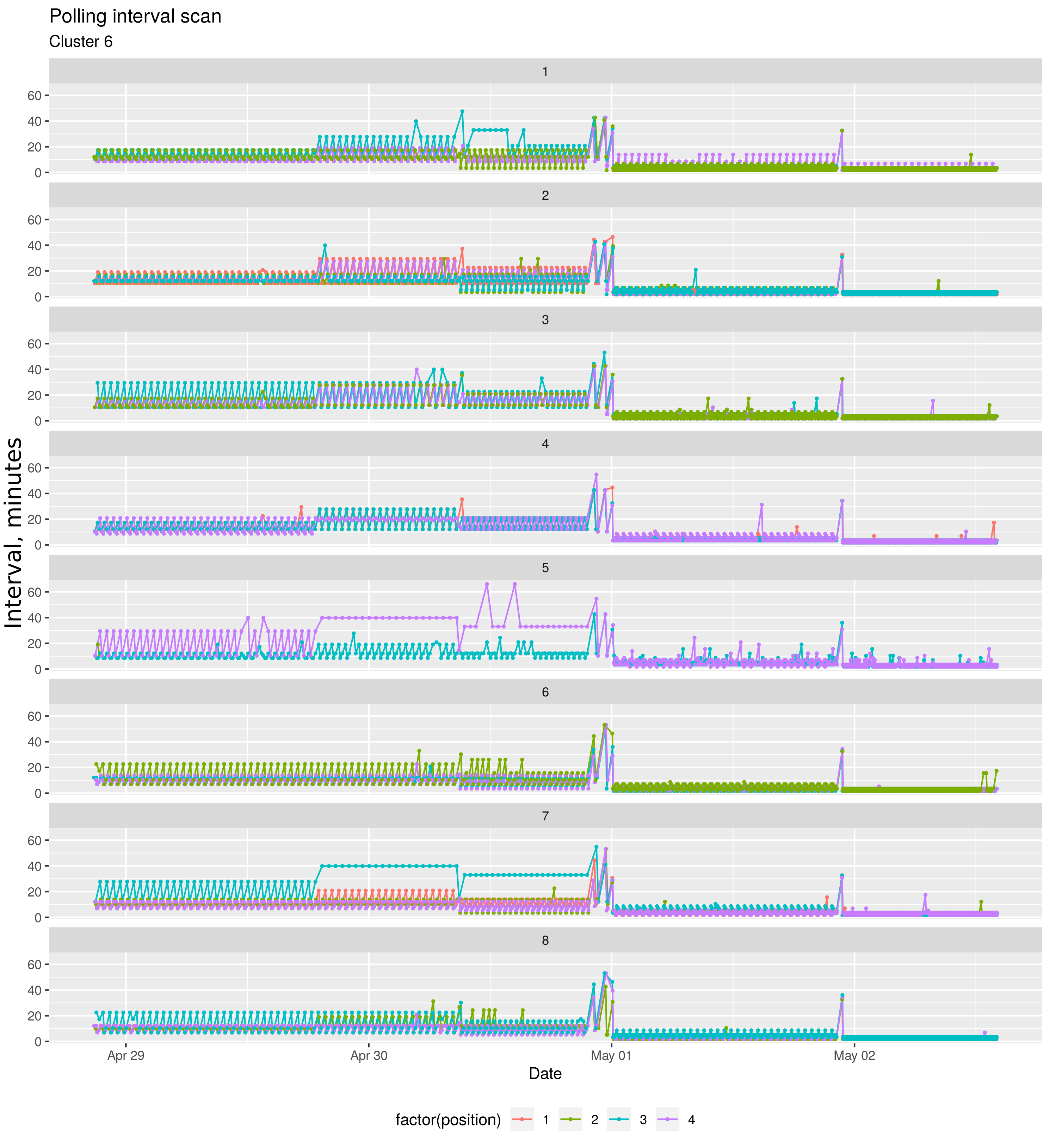}
      \caption{Polling intervals for beacons on cluster 6}
      \label{fig:ierr_2}
    \end{subfigure}
    \caption{Beacon polling intervals from April 29th 2020 to May 3d 2020. The period from April 29th to May 1st corresponds to a purely non-optimized polling. The period from May 1st to May 2nd corresponds to an old optimized polling procedure. Finally, the period from May 2nd onwards corresponds to a new polling algorithm. Each facet corresponds to a different string of the particular cluster.}
    \label{fig:polling}
  \end{figure}

  With the addition of each new cluster, at least 32 new AMs are added to the APS, including at least 4 floor nodes.
  This translates into an increase in the polling interval, which in turn raises beacon positioning uncertainty.
  With the the average beacon speed of 0.5 cm/s (see \cite{icrc2019pos}), a beacon can drift over 3 meters from the previous measurement before a new measurement is complete.
  During the active drift period, this distance can reach over 15 meters.
  In May 2020, a new polling algorithm was introduced, allowing for a parallel beacon polling and reducing polling period by a factor of 2 to 3. 
  Figure \ref{fig:polling} demonstrates the dramatic reduction in polling times following the introduction of the new polling procedure.

  \subsection{Long term beacon drift}
  
\begin{figure}[h]
    \centering
    \begin{subfigure}[t]{0.45\textwidth}
      \centering
      \includegraphics[width=\textwidth]{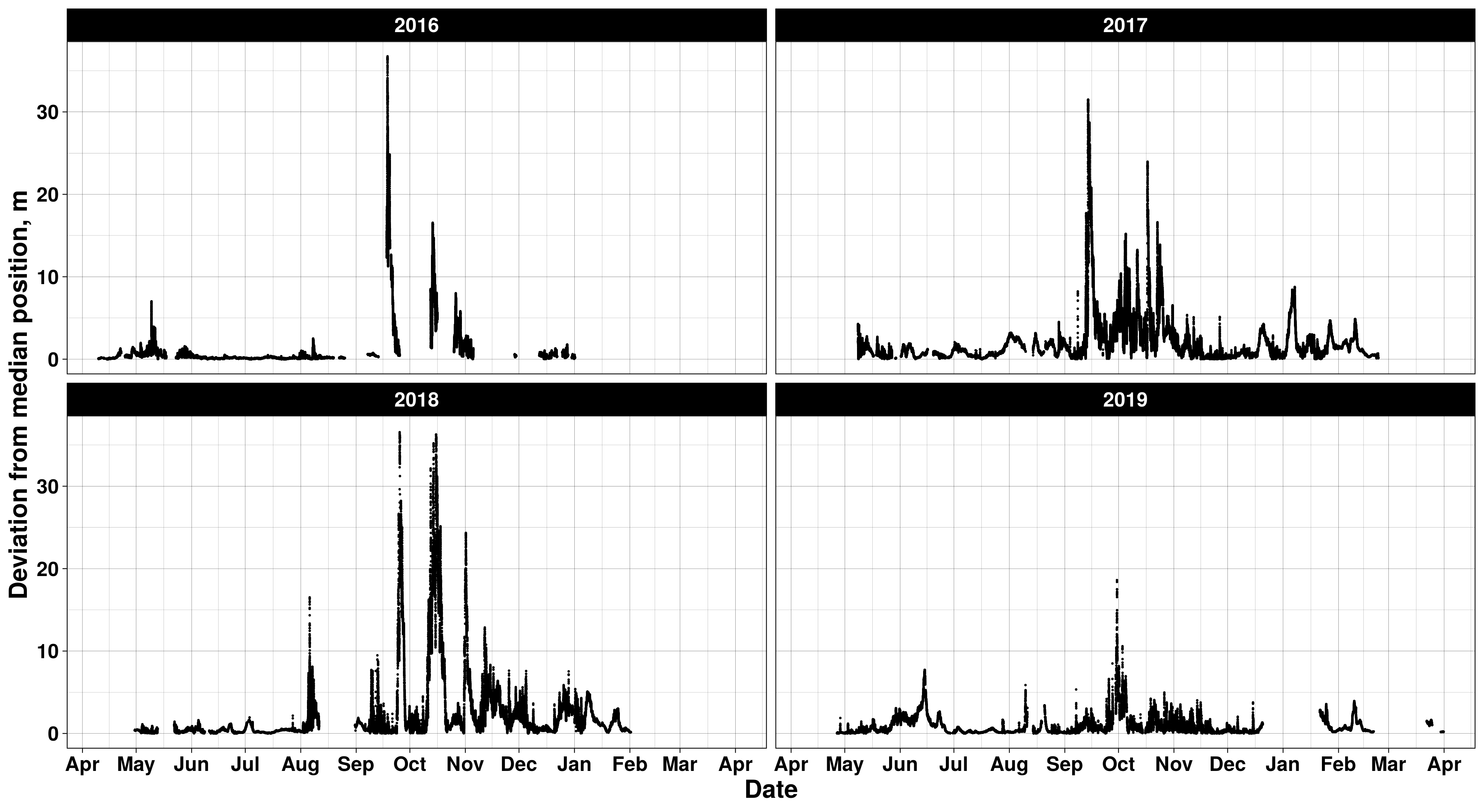}
      \caption{Deviation from stable position over time.}
    \end{subfigure}
    \hfill
    \begin{subfigure}[t]{0.45\textwidth}
      \centering
      \includegraphics[width=\textwidth]{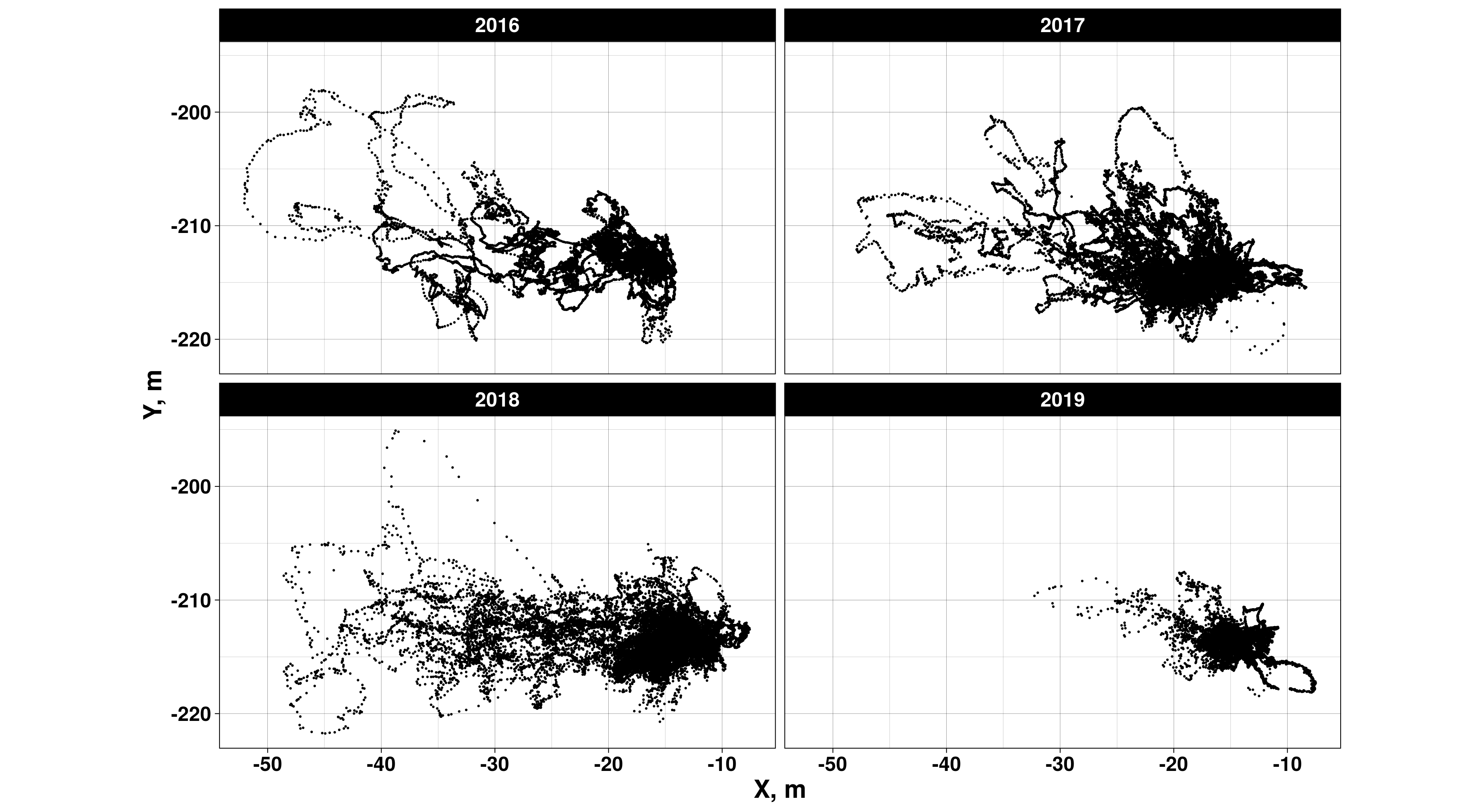}
      \caption{Total planar coordinates}
    \end{subfigure}
    \caption{Long term behaviour of beacon \#48, installed at the depth of 920 m over 4 seasons.}
    \label{fig:long}
  \end{figure}

  Long term beacon drift (see Figure \ref{fig:long}) exhibits similar patterns from season to season.
  Deviation from median position for a particular beacon is typically within 10 meters until late August.
  Starting from late August the GVD may occasionally enter the active drift periods during which beacons can move as far as 50 meters beyond their stable positions. 
  High dirft periods are disjoint and can last up to a week each, totalling nearly a month each season.
  String movement during active drift periods is correlated between multiple clusters (see \cite{icrc2019pos}) and are dominated by the movement along the shore.

  \subsection{Short term beacon drift}

\begin{figure}[h]
    \centering
    \begin{subfigure}[t]{0.55\textwidth}
      \centering
      \includegraphics[width=\textwidth]{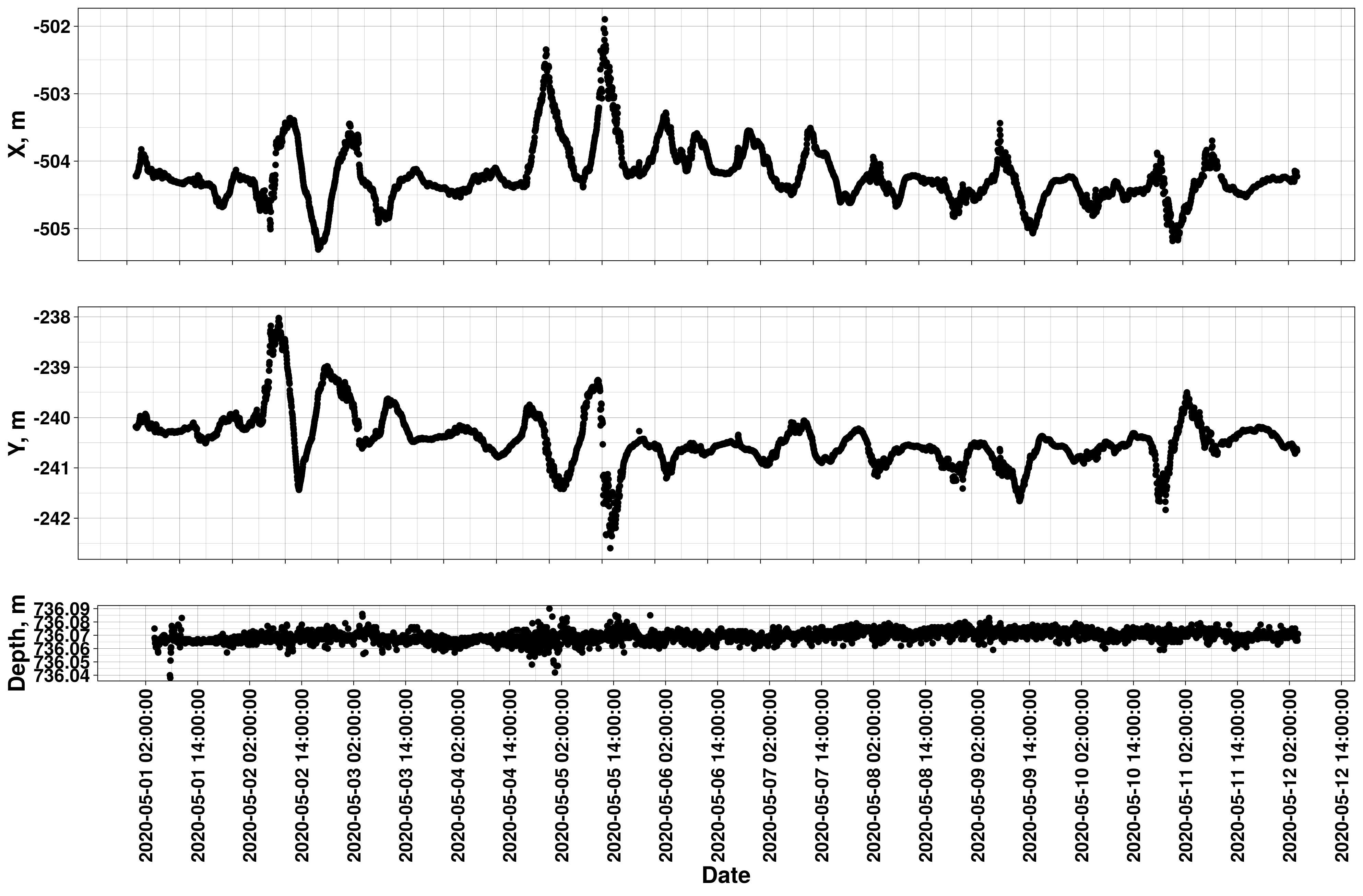}
      \caption{Coordinate scans}
      \label{fig:short-scan}
    \end{subfigure}
    \hfill
    \begin{subfigure}[t]{0.35\textwidth}
      \centering
      \includegraphics[width=\textwidth]{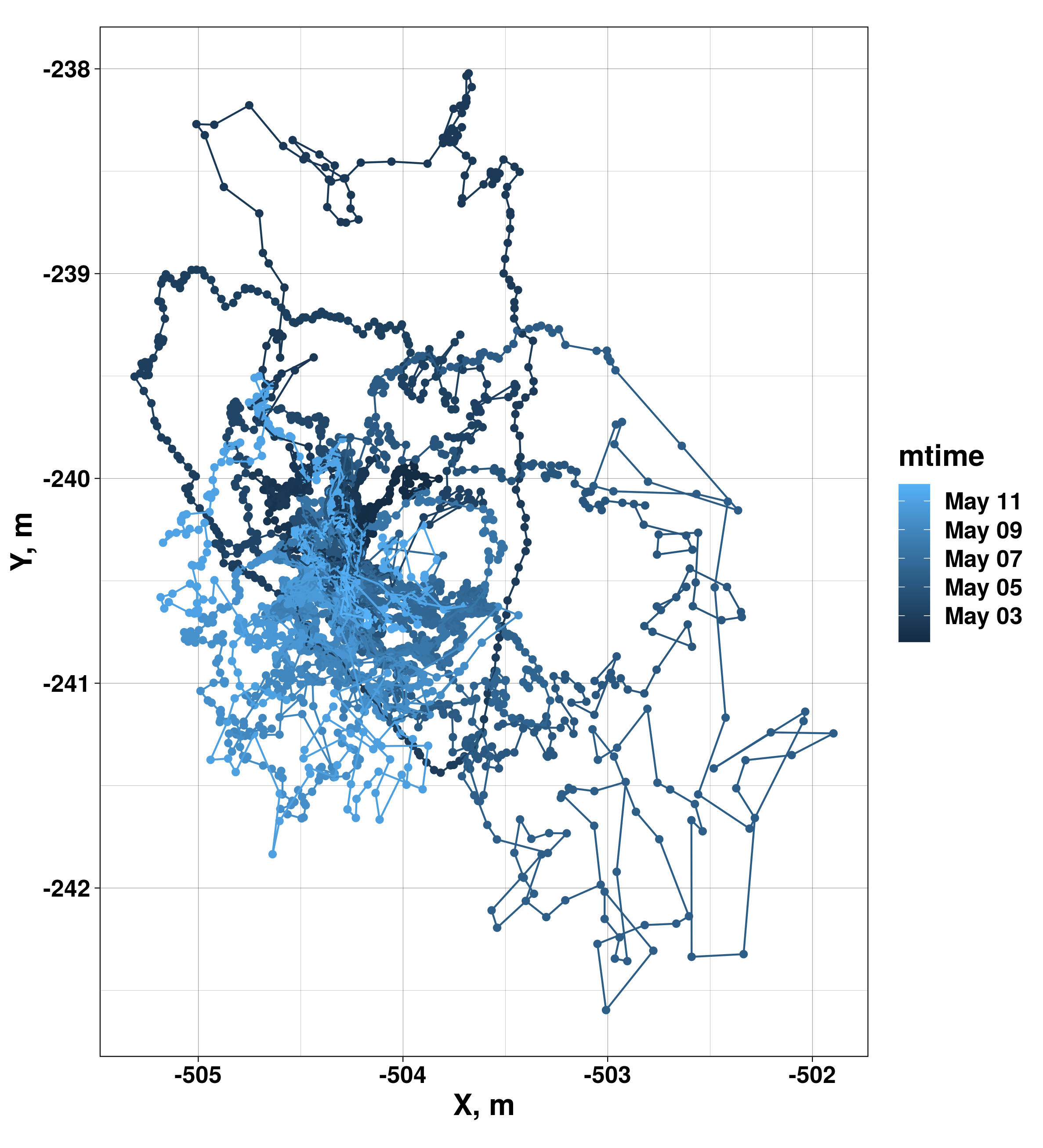}
      \caption{Planar coordinates}
    \end{subfigure}
    \caption{Beacon \#306 drift between May 1 and May 12 2020, during a regular drift period.}
    \label{fig:short}
  \end{figure}

  Short term beacon drift for a shallow beacon for a regular drift period are shown on Figure \ref{fig:short}.
  As can be seen on Figure \ref{fig:short-scan}, most of beacon movement occurs in the XY plane, while the depth variation is minimal.
  The scale of beacon deviation from a stable position in this period is typically several meters.
 
  \subsection{Positioning precision}

\begin{figure}[h]
    \centering
    \begin{subfigure}[t]{0.45\textwidth}
      \centering
      \includegraphics[width=\textwidth]{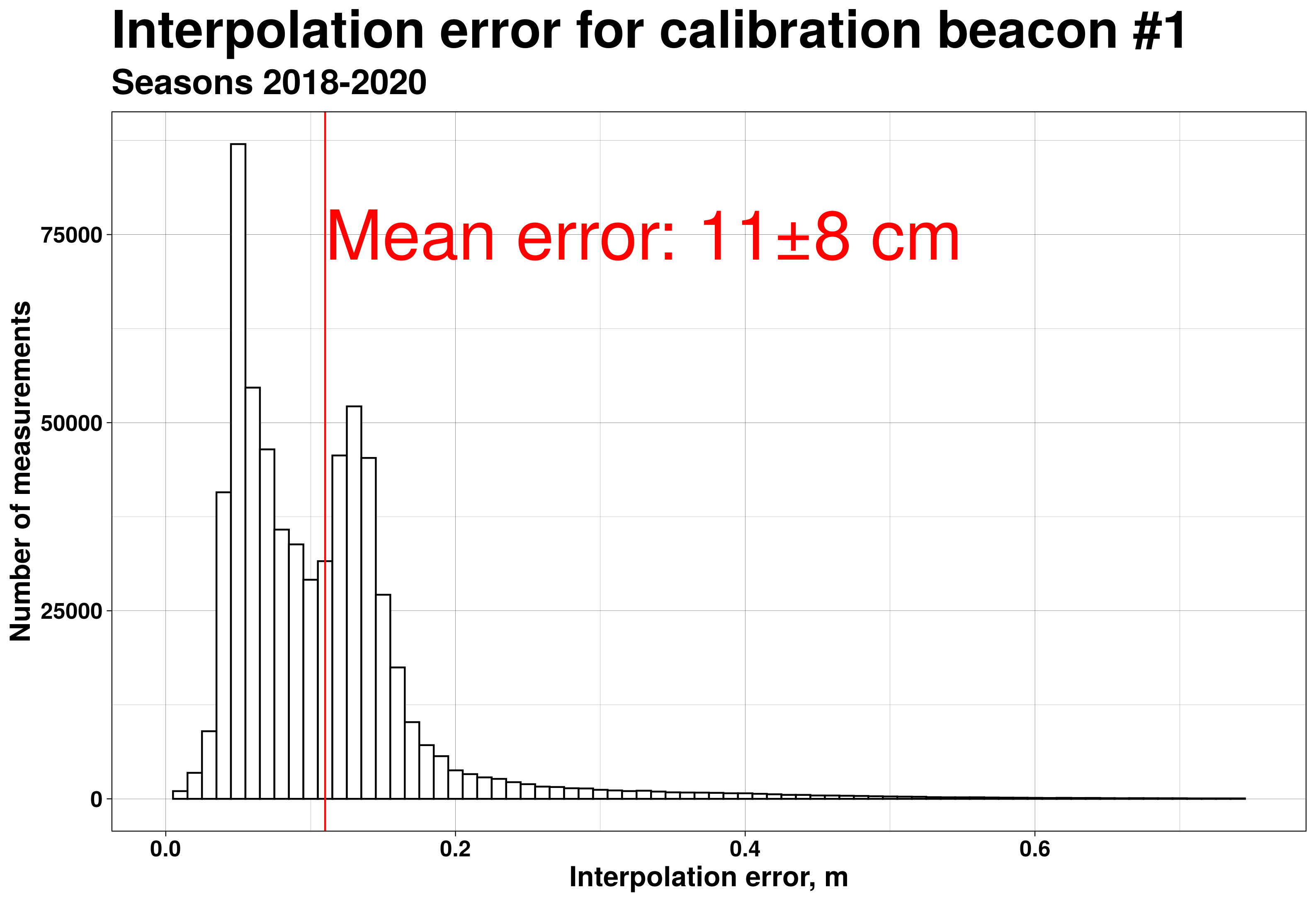}
      \caption{Interpolation error for calibration beacon 1}
      \label{fig:ierr_1}
    \end{subfigure}
    \hfill
    \begin{subfigure}[t]{0.45\textwidth}
      \centering
      \includegraphics[width=\textwidth]{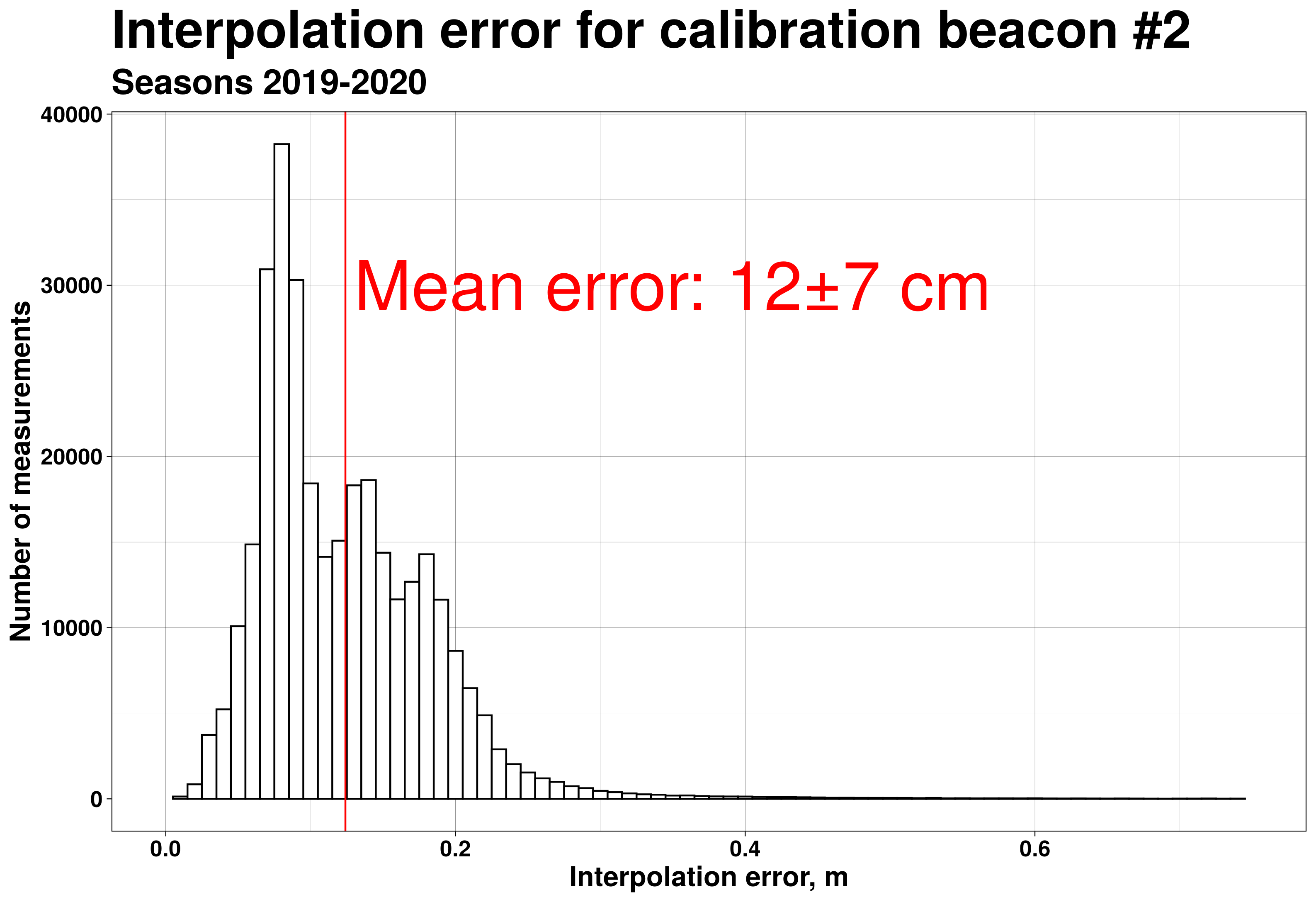}
      \caption{Interpolation error for calibration beacon 2}
      \label{fig:ierr_2}
    \end{subfigure}
    \caption{Interpolation errors for calibration beacons.}
    \label{fig:ierr}
  \end{figure}

  OM positions are interpolated from beacon coordinates using a piecewise-linear string model.
  In this approach, the interpolation error will vary with the distance between OM and interpolation beacons, string curvature, and beacon mobility, which is decreases with depth and varies throughout the season.
  The biggest interpolation error is expected for OMs installed in the middle between beacons 3 and 4.
  These beacons are the most mobile due to their relatively shallow position. 
  The string segment between beacons 3 and 4 also exhibits greater curvature compared to lower string segments, compounding linear interpolation error.
  In order to obtain experimental bounds on this worst-case error, two calibration beacons were installed in the middle of the string segment between regular beacons 3 and 4.
  Calibration beacon \#1 was installed on the central string of cluster 3 in 2018, and calibration beacon \#2 was installed on a peripheral string of cluster 5 in 2019.
  Calibration beacon coordinates were then linearly interpolated from the coordinates of beacons 3 and 4. 
  The distance between calibration beacon positions obtained with linear interpolation and acoustic trilateration then provides an upper bound on OM positioning error.
  The distributions of these differences are provided on Figure \ref{fig:ierr}.
  As can be seen from this analysis, the mean upper bound on OM positioning error is below 15 cm (note that the photocathode diameter is 25 cm).
  This bound is possible because GVD strings are largely nearly vertical.
  This is consistent with the results of a previous study, done over the course of season 2018 \cite{icrc2019pos} and corresponds to a subnanosecond time calibration error.
  While the mean error over three years is comparable to the positioning error in similar detectors \cite{riccobene}, the immediate interpolation error during active drift periods can reach 0.5 meters due to high beacon mobility and string curvature.
  This can only be ameliorated by applying beacon coordinates to a physical, rather then piecewise-linear string model.

  \section{Conclusion}
  The Baikal-GVD acoustic positioning system is currently in operation.
  Following an improvement in the polling algorithm in May 2020, the median polling interval in 2021 is 180 seconds.
  Short-term and long-term beacon drifts were presented.
  The mean positioning error has been estimated with three years of data to be below 15 cm, which corresponds to a subnanosecond hit time uncertainty and is in agreement with previous work.

  \section{Acknowledgements}
  The work was partially supported by RFBR grant 20-02-00400.
  The CTU group acknowledges the support by European Regional Development Fund-Project No. CZ.02.1.01/0.0/0.0/16\_019/0000766.
  We also acknowledge the technical support of JINR staff for the computing facilities (JINR cloud).

  %\begin{thebibliography}{99}
  \bibliographystyle{unsrt}
  \bibliography{procbib}
  %\end{thebibliography}

  \end{document}